\documentclass[12pt,preprint]{aastex}

\shortauthors{Sil'chenko and Moiseev}
\shorttitle{Nuclear rings in unbarred galaxies}

\received{2005 May 17}

\begin{document}

\title{Nature of nuclear rings in unbarred galaxies: NGC 7742 and NGC 7217}
\slugcomment{Partly based on observations collected with the 6m
telescope at the Special Astrophysical Observatory (SAO) of the
Russian Academy of Sciences (RAS).}

\author{O. K. Sil'chenko\altaffilmark{1}}
\affil{Sternberg Astronomical Institute, Moscow, 119992 Russia\\
       Isaac Newton Institute, Chile, Moscow Branch\\
     Electronic mail: olga@sai.msu.su}
\altaffiltext{1}{Guest Investigator of the UK Astronomy Data Centre}

\and

\author{A. V. Moiseev}
\affil{Special Astrophysical Observatory, Nizhnij Arkhyz,
    369167 Russia\\
    Electronic mail: moisav@sao.ru}

\begin{abstract}

We have studied the unbarred Sb galaxy with a nuclear star-forming
ring, NGC 7742, by means of 2D spectroscopy, long-slit spectroscopy,
and imaging, and have compared the results with the properties of another
galaxy of this type, NGC 7217, which is studied by us earlier. Both galaxies
have many peculiar features in common: each has two global exponential
stellar disks with different scalelengths, each possesses a circumnuclear
inclined gaseous disk with a radius of 300 pc, and each has a global
counterrotating subsystem, gaseous one in NGC 7742 and stellar one in NGC 7217.
We suggest that past minor merger is the probable cause of all these
peculiarities, including appearance of the nuclear star-forming rings without
global bars; the rings might be produced as resonance features by tidally
induced oval distortions of the global stellar disks.

\end{abstract}

\keywords{galaxies: individual (NGC7742) ---
galaxies: individual (NGC7217) --- galaxies: evolution ---
galaxies: structure}

\section{Introduction}

Nuclear rings, looking prominent features due to their intense star
formation, are found mostly in barred galaxies so they are commonly
treated as linked to inner Lindblad resonances where all radial gas
inflows are slowed down and where gas is accumulated \citep{butcr93,shlosrings}.
However there are several cases of spectacular nuclear rings in unbarred
galaxies, such as those in NGC 278, NGC 7217, NGC 7702, NGC 7742.
Most of these galaxies are seen face-on, so the conclusion about
the bar absence is quite safe in them. Suggestions about the nature
of nuclear rings in unbarred galaxies includes: resonance effects
produced by weakly triaxial potential \citep{jungpal,bu95};
resonance effects produced by a past bar which is now dissolved \citep{ath};
viscous gas accretion produced by rotation velocity shear in the global disk
and its accumulation at a stagnation point at the turnover radius of the
rotation curve \citep{wegmc2}; finally, minor merger \citep{n278}.
Among these hypotheses, the last provides more opportunities
to explain various combinations of observational facts. Indeed,
\citet{athcoll} have shown that vertical central impact
of a small satellite whose mass is about 10\%\ of the host mass
should produce a nuclear stellar ring which is morphologically
indistinguishable from a resonance ring. On the other hand, if the
merged satellite orbit was close to the main galaxy disk plane, their
gravitational interaction might produce an oval disk distortion which
could in its turn create a resonance nuclear ring. However, certain
combinations of predictions are provided by each theoretical model,
and by collecting more various observational data, both morphological
and kinematical, for every galaxy in question, we would be able at last
to restrict possible mechanisms of nuclear ring generation in any
particular case.

In this paper we will consider NGC 7742 and NGC 7217; both galaxies have
prominent nuclear star-formation rings with a radius of some $10\arcsec$.
We will attempt to find any general features which may be connected to the
nuclear ring origin. As for the latter galaxy, now we are undertaking our
third approach to its study. Earlier we have found a circumnuclear gas polar
ring and two exponential stellar disks with different scalelengths in it
\citep{we97c,we2000}. Also, NGC 7217 is known to possess two counterrotating
stellar subsystems \citep{mk94,we2000}. Recently the SAURON team has also
found a gas-stars counterrotation in the center of NGC 7742 \citep{sau2},
so this fact promises interesting speculations. Both galaxies are
moderate-luminosity unbarred spirals of Sb-type.

\section{Observations and data used}

New observational data which we intend to analyse in this work
concerns mainly NGC 7742:
panoramic spectral data for NGC 7742 have been obtained with
the scanning Fabry-Perot Interferometer (IFP) of the
6m telescope of the Special Astrophysical Observatory of the Russian Academy
of Sciences (SAO RAS) and with two integral-field spectrographs,
the fiber-lens Multi-Pupil Fiber Spectrograph (MPFS) at the 6m telescope
of the SAO RAS and the international Tigre-mode SAURON at the 4.2m William
Herschel Telescope at La Palma (see Table~1). Our 2D spectroscopic data for
NGC 7217 have been described in detail earlier: the Fabry-Perot
data -- by \citet{we97c} and the MPFS data -- by \citet{we2000}.

\subsection{2D spectroscopy with the MPFS}

The present modification of the MPFS of the 6m telescope works
at the prime focus from the summer of 1998 \citep{mpfsref}; see also
http://www.sao.ru/hq/lsfvo/devices/mpfs/).
NGC 7742 was observed with the MPFS several times during 2001--2003.
MPFS is a fiber-lens system: densely packed square microlenses placed
in the focal plane of the telescope create a set of $16\times 15$
micropupils, or $16\times 16$ in 2003,
and the fibers after them transmit the light from the square
elements of the galaxy image to the slit of the spectrograph together
with 16 (17) additional fibers that transmit the sky background light taken
at a distance of $4^{\prime}$ from the galaxy, so the sky spectra
are obtained together with those of the target. The size of one spatial
element is approximately
$1\arcsec \times 1\arcsec$; a CCD TK $1024 \times 1024$ and during the
latest run of October 2003 -- a CCD EEV 42-40 $2048 \times 2048$ were used.
The spectral resolution was about 4~\AA\ varying by about 20\%\ over
the field of view. The wavelength calibration was done with
a He-Ne-Ar lamp before and after the galaxy exposures;
the internal accuracy of linearization was typically 0.25~\AA\ in the
green and 0.1~\AA\ in the red. Also we checked the accuracy of
the wavelength calibration and the absence of a systematic velocity shift
by measuring strong emission lines of the night sky, [\ion{O}{1}]$\lambda$5577
 and [\ion{O}{1}]$\lambda$6300. We obtained the MPFS data in two spectral
ranges, green, 4300--5600~\AA, and red, 5900--7200~\AA.
The green spectra were used to obtain the line-of-sight velocity field for
the stellar component and a map of the stellar velocity dispersion
by their cross-correlation with spectra of some template stars,
usually of G8III--K1III spectral type. The red spectral range which
contains strong emission lines H$\alpha$ and [\ion{N}{2}]$\lambda$6583
is appropriate to derive line-of-sight velocity fields of the ionized gas.

\subsection{2D spectroscopy with SAURON}

The other 2D spectrograph which data we use is a rather new
instrument, SAURON, installed at the 4.2m William Herschel Telescope (WHT)
on La Palma -- for its detailed description see \citet{betal01}
and for some preliminary scientific results see \citet{sau2}.
We have taken the data for both our galaxies, NGC 7217 and NGC 7742
observed in October 1999, from the open ING Archive of the UK Astronomy
Data Centre. If to give a brief description, the field of view of this
instrument is $41\arcsec \times 33\arcsec$
with a spatial element size of $0\farcs 94 \times 0\farcs 94$.
The sky background is taken less than 2 arcminutes from the center of the
galaxy and is exposed simultaneously with the target. The spectral
range is fixed as of 4800-5400~\AA, the spectral resolution is about 4~\AA,
also varying over the field of view.
The comparison spectrum is that of pure neon, and to made the
linearization we fit a polynomial
of the 2nd order with an accuracy of 0.07~\AA.

\begin{table*}
\scriptsize
\caption[ ] {Integral-field spectroscopy of the galaxies studied}
\begin{flushleft}
\begin{tabular}{lllllcc}
\hline\noalign{\smallskip}
Date & Galaxy & Exposure & Configuration & Field
& Spectral range & Seeing \\
\hline\noalign{\smallskip}
14 Oct 99 & NGC~7217, Pos.1 & 120 min & WHT/SAURON+CCD $2k\times 4k$ &
$33\arcsec\times 41\arcsec$ & 4800-5400~\AA\ & $1\farcs 4$ \\
14 Oct 99 & NGC~7217, Pos.2 & 120 min & WHT/SAURON+CCD $2k\times 4k$ &
$33\arcsec\times 41\arcsec$ & 4800-5400~\AA\ & $1\farcs 4$ \\
13 Oct 99 & NGC~7742 & 120 min & WHT/SAURON+CCD $2k\times 4k$ &
$33\arcsec\times 41\arcsec$ & 4800-5400~\AA\ & $1\farcs 1$ \\
22 Sep 01 & NGC~7742 & 45 min & 6m/MPFS+CCD $1024 \times 1024$ &
$16\arcsec \times 15\arcsec $ & 4200-5600~\AA\ & $2\farcs 1$ \\
2 Oct 03 & NGC~7742 & 20 min & 6m/MPFS+CCD $2048 \times 2048 $ &
$16\arcsec \times 16\arcsec $ & 5800-7200~\AA\ & $2\farcs 0$ \\
\hline
\end{tabular}
\end{flushleft}
\end{table*}

\subsection{2D spectroscopy with the IFP}

In November 2003, NGC 7742 has been observed with the scanning Fabry-Perot
Interferometer (IFP) of the 6m telescope installed at the prime focus within
the focal reducer SCORPIO \citep{scorpref}; see also
http://www.sao.ru/hq/moisav/scorpio/scorpio.html.
The total number of 32 spectral channels were exposed, each during 3 minutes,
providing the spectral resolution of 2.5~\AA. The seeing was
$1\farcs 7 -2\farcs 1$; the spatial binning used was $0\farcs 7$ per pixel, and
the full field of view obtained was $6\arcmin \times 6\arcmin$. The narrow
filter centered on the spectral region around redshifted H$\alpha$ and
[\ion{N}{2}]$\lambda$6583 emission lines was used. The velocity field of
the ionized gas obtained by measuring the H$\alpha$ is more precise and
extended whereas the measurements of [\ion{N}{2}] allow to probe the very
center of the galaxy where the H$\alpha$ emission is strongly contaminated by
the absorption line.

As we discuss below, both velocity fields give consistently the orientation
parameters of the gas disk: the kinematical major axis, or the line-of-nodes,
at $PA=128\degr$ and inclination of $i=9\degr$.

\subsection{Long-slit spectroscopy of NGC 7742}

To supplement our 2D spectroscopy by additional data, we have retrieved
some long-slit data for NGC 7742 from the ING Archive: the galaxy was
observed in November 1997 with the two-armed ISIS spectrograph of the
4.2m William Herschel Telescope. However, the quality of these data
seems to be insufficient: evidently, the spectral focus was not checked
promptly, and the spectral resolution was bad. So we have only measured
baricenters of the most prominent emission lines in the long-slit
cross-section of $PA=160\degr$ to determine line-of-sight velocities
of the ionized gas near the center of the galaxy; either gas velocity
dispersion nor stellar kinematics are not probed. This direction of
the slit, $PA=160\degr$, is not very close to the kinematical major axis
of the gas. So to obtain more conclusive data, in November 2004 we have
observed NGC 7742 at the 6m telescope with the focal reducer SCORPIO in the
long-slit mode with a large spectral range of 5700-7200~\AA\ and a spectral
resolution of about 5~\AA. The seeing was about $1\farcs 5$.
The slit of $1\arcsec$ width was aligned with
the kinematical major axis at $PA=128\degr$. Here we analysed both the
emission lines and the absorption line of \ion{Na}{1}D to probe the
kinematics of the
stellar component. The K-giants HD 4744 and 20893 were observed the same
night at the same mode; their spectra were used for cross-correlation
with the spectra of the galaxy.

\subsection{Imaging data}

The same night, on 5th of November, 2004, we obtained a rather deep
V-image of NGC 7742 during 240 s with the focal reducer SCORPIO in the
imaging mode (pixel scale was $0\farcs 36$ and the seeing was $1\farcs 7$).
Besides, we have retrieved large-scale B- and I-filter images for this galaxy
from the ING Archive (the images are obtained with the one-meter Jacobus
Kapteyn Telescope, the scale is $0\farcs 33$ per pixel and the seeing was
$1\arcsec$) as well as small-scale HST/NICMOS2 images (with the scale of
$0\farcs 075$ and the spatial resolution of $0\farcs 2$) and HST/WFPC2
images (with the scale of $0\farcs 1$ and the spatial resolution of
$0\farcs 2$) from the HST Archive. To check the large-scale structure
of the galaxies in the NIR, we have used the 2MASS images taken from the
NASA/IPAC Extragalactic Database.

The data have been mostly analysed by using the software
produced by Dr. V.V. Vlasyuk of the Special Astrophysical Observatory
\citep{vlas}; only primary reduction of the data obtained with the MPFS
and SCORPIO (images and long slit) was done in IDL with various pieces
of software created by one of us (A.V.M.) and by Prof. V. L. Afanasiev.
The V-image of NGC 7742 obtained with the reducer SCORPIO has been
calibrated into the standard Johnson system by using the single
photoelectric aperture measurement by \citet{keel78}.
The data observed with the IFP were reduced with the
IDL-based software described by \citet{ifpref}. Also the ADHOC
package\footnote{ADHOC software is written by J. Boulesteix
(Observatoire de Marseille).
See \texttt{http://www-obs.cnrs-mrs.fr/ADHOC/adhoc.html}}
was involved to smooth the ``data cubes''. The monochromatic images
and velocity fields of the emission lines H$\alpha$ and
[\ion{N}{2}]$\lambda$6583
were constructed by means of fitting the IFP spectra with Gaussians.
We analysed two kinds of data: one with the original spatial resolution
($2\farcs 2$) and the other smoothed by a gaussian filter with
FWHM of $2\times 2$ elements (the spatial resolution of $2\farcs 7$).
The results are mainly the same, but the last data are better
for the low-brightness regions.

\section{Stellar and gaseous kinematics of NGC 7742}

\citet{sau2} having presented the SAURON 2D velocity fields
both for stars and for the ionized gas in NGC 7742 -- the latter obtained
from their sophisticated measurement of the weak [\ion{O}{3}]$\lambda$5007
emission line -- have claimed an appearance of strict counterrotation
of the stars versus the gas. However they have shown only the very central
parts of the velocity fields within the nuclear ring ($R\approx 10\arcsec$),
so it has remained unclear if we deal with a global counterrotation or
with a compact circumnuclear counterrotating gaseous disk.
Figure~\ref{ifp7742} presents
our large-scale Fabry-Perot observations of NGC 7742 in the H$\alpha$
and [\ion{N}{2}] emission lines. In the upper two rows we give the distributions
of the emission-line intensities (left) and the velocity fields (right),
the bottom right plot presents the results of the velocity field analysis
made by a tilted-ring method \citep{begeman}. One can see that despite
the face-on view of the galactic disk, the line-of-sight velocity field
of the ionized gas demonstrates a quite regular rotation up to the
border of the noticeable H$\alpha$ emission at $R=30\arcsec -40\arcsec$,
with the visible amplitude of line-of-sight velocity variations of
about 40 km/s.
By fixing the kinematic center position that coincides reasonably well
with the photometric center and also the systemic velocity and by assuming
the same orientation angles, line-of-nodes position angle $PA_0$ and
inclination $i$, for the whole gaseous disk, we obtain rather sure
estimates of the disk orientation parameters: $i=9\degr \pm 4\degr$ and
$PA_{0,kin,gas}=128\degr \pm 1\degr$. With these parameters of the disk
orientation, the azimuthally averaged circular rotation velocity can be
estimated as 220-230 km/s within the nuclear ring radius, $R\le 10\arcsec$;
outside it decreases smoothly to $\sim 150$ km/s at $R\approx 40\arcsec$.
At the outer edge of the nuclear disk we note a drop of the rotation velocity
by $\sim 30$ km/s; another drop, more prominent in the H$\alpha$ velocity
field than in the [\ion{N}{2}] velocity field, can be detected at the radii of
$25\arcsec -28\arcsec$. As we shall show below, the latter radius is also
distinguished photometrically. So we may conclude that the sense of the
gas rotation (which is opposite to the stars rotation in the center) remains
unchanged up to the large radii. We can even expand the spatial range, over
which the conclusion is valid, beyond the borders of H$\alpha$ emission.
\citet{knapp} measured an emission line of the neutral hydrogen,
$\lambda$21 cm, in several positions near NGC 7742. Though their spacing and
half-beam resolution was rather rough, of about $2^{\prime}$, they detected
a noticeable rotation `in an east-west direction', the eastern side of the
HI disk being receding. So we conclude that NGC 7742 possesses the large
gaseous disk which rotates regularly so that its eastern side is receding;
the line of nodes of this disk is at $PA=128\degr$, and the inclination can
be determined kinematically as $i\approx 9\degr$.

Figure~\ref{mpfs7742} presents the circumnuclear velocity field
of the ionized gas which
we have obtained with the MPFS by measuring the strongest emission line
of this region, [\ion{N}{2}]$\lambda$6583. These data compliment
the large-scale gas velocity field obtained with the IFP. Over this
velocity field also, we see a regular rotation, with the visible amplitude
of the line-of-sight velocity variations of $\pm 60$ km/s, and its kinematical
major axis at radii larger than
$2\arcsec$ can be  determined quite certainly as $PA_0=128\degr$, being
completely consistent with the line of nodes of the global gaseous disk.
However, if we apply the inclination of $9\degr$ found for the global gas disk
to the circumnuclear velocity field we would obtain a formal value of the
circular rotation velocity of 400 km/s at $R\approx 3\arcsec$. This seems
improbable for the galaxy of such moderate luminosity,
inconsistent with the Tully-Fisher relation; moreover, the central stellar
velocity dispersion in NGC 7742 estimated by us both with the MPFS and the
SAURON data is less than 80 km/s, so there are no any signs of huge mass
concentration in the nucleus of this galaxy. We should rather conclude that
the inclination of $9\degr$ is not valid for the very central part of the
gaseous disk, $R<4\arcsec$, and that the disk begins to warp when approaching
the nucleus. The SAURON data
were obtained under better seeing conditions than ours, and in the recently
delivered Ph.D. Thesis of Kambiz Fathi \citep{fathithes} the gas velocity
field of NGC 7742 reveals a turn of its kinematical major axis
by $90\degr$ at $R<2\arcsec$. The comment of Dr. Fathi is that we see radial
gas motions. But if the gaseous disk remains to be nearly face-on around the
nucleus, we would not see any noticeable projection of radial velocities
confined within the disk
onto the line of sight. Indeed, if we accept the inclination of $9\degr$
obtained for the whole gaseous disk for the very center of NGC 7742,
the Fathi's results would imply an amplitude of the possible radial motions
exceeding twice the rotation velocities -- namely, of $\pm \sim 400$ km/s.
We do not see any reason to suspect such
supersonic radial gas flows in the morphologically regular galaxy
with the nuclear activity of a rather weak LINER/transition type.
More probably, we see here an inclined circumnuclear disk similar
to those found in some spiral galaxies, and in
particular in NGC 7217 \citep{we97c,we2000}.

As for the stellar rotation, it does not seem to be so fast as that of the
gas and is mostly confined to the very inner, $R<3\arcsec$, region of the
galaxy. Since the seeing conditions during our MPFS observations of NGC 7742
were not good enough to resolve this small region properly, the measured
amplitude of the stellar line-of-sight velocity variations is dropped due to
spatial smoothing, and the orientation of the kinematical major axis cannot
be determined properly from our data. In Fig.~\ref{sau7742} we show our analysis
of the SAURON velocity field for the stars in the center of NGC 7742. We have
obtained $PA_{0,kin,*}=335\degr$ for the stellar component within $R=6\arcsec$,
in some disagreement with $PA_{0,*}=320\degr$ found by Fathi; however,
the possible error may be as large as $10\degr$. If again we formally fix
the inclination of the rotation plane of the stars at the value obtained
for the outer gaseous disk, $i=9\degr$, the peak rotation velocity achieved
at $R=1\arcsec$ would be $v_{rot}\approx 250$ km/s. Farther from the nucleus
it drops to zero at the radius of the ring, and beyond the ring it rises
marginally. Due to low signal-to-noise ratio we are not sure with our results
at $R>10\arcsec$, and to check if the stars continue to counterrotate the
gas outside the ring radius we appeal to the long-slit data.

Figure~\ref{ls7742} presents long-slit velocity profiles for the stars and
ionized gas:
SCORPIO data along the kinematical major axis $PA=128\degr$ and WHT/ISIS
data at $PA=160\degr$. Because of the low surface brightness of NGC 7742
at $R>10\arcsec$, the measurements of the stellar velocities
(Fig.~\ref{ls7742}{\it a}) are
not very extended and are not very precise; however some qualitative
conclusions can be made. The sense of rotation of the stellar component
observed in the center persists up to $R\approx 25\arcsec$ at least; but at
$R\approx 10\arcsec$ -- at the radius of the ring -- we see strong stellar
velocity variations, such that the line-of-sight velocities of the stars at
this radius coincide exactly with those of the ionized gas. We may suggest
that violent star formation in the ring has already produced a substantial
stellar population, including stars of F-G-K type, so that their rotation
coupled with their parent gas contributes significantly into the integrated
LOSVD of the stars at this radius.

The long-slit gas velocity profiles (Figs.~\ref{ls7742}{\it b} and {\it c})
demonstrate different character with respect to the rather smooth rotation
curve obtained by azimuthal averaging of the 2D IFP gas velocity field
(Fig.~\ref{ifp7742}). They
'oscillate' by 70-80 km/s with a characteristics radial period of
$\sim 10\arcsec$, and the locations of the velocity maxima and minima differ
at $PA=128\degr$ and at $PA=160\degr$. We think that these velocity variations
do not relate to regular rotation; they resemble vertical small-scale
oscillations of a tidally perturbed gaseous disk.

Another peculiarity of the long-slit gas velocity profiles which we have
however expected basing on our 2D MPFS data is the very steep central
velocity gradient and the decoupled fast gas rotation within $R=3\arcsec$;
it is confirmed both by the SCORPIO and ISIS data. Moreover, the [\ion{N}{2}]
emission line measurements at the approaching branch of the velocity profile
give even underestimated values of the rotation velocity -- they deviate
toward the systemic velocity not only with respect to the H$\alpha$
measurements which may be affected by underlying
absorption lines, but also with respect to the [\ion{S}{2}] emission line
measurements. Similar differences between
velocity estimates made with different emission lines had been detected more
than once in the centers of other spiral galaxies\citep{mrk744,ourls2} and
might be explained if the H$\alpha$ and
[\ion{S}{2}] emission lines relate to regularly rotating gas ionized
by OB stars and the [\ion{N}{2}] emission lines are formed mostly in shock
wave sites where the ionized gas decelerates. Enormous visible rotation of
the ionized gas within $R=3\arcsec$ revealed by the long-slit data gives
strong evidence for the highly-inclined orientation of the gas
rotation plane in the very center of NGC 7742 ($i_{gas} >35\degr$),
as opposite to the nearly face-on orientation of the global gaseous disk.

\section{Global structure of NGC 7742}

The morphological type of NGC 7742 is
SA(r)b, and taking into account the face-on orientation of the global disk,
the galaxy looks indeed quite round and axisymmetric, except the very
central part (Fig.~\ref{ifp7742}). However, the rather early morphological
type, Sb, deduced perhaps from the appearance of tightly wound, faint spiral
arms, is not supported by the very  low stellar velocity dispersion in the
center, $\le 80$ km/s, implying the absence of a large bulge that is obliged
to be a dynamically hot stellar subsystem by definition.

The V-band image obtained with the SCORPIO appears to be very deep: our
surface brightness measurements reaches the radius almost twice that of 25th
B-magnitude. Figure~\ref{iso7742} presents the results of isophotal analysis
of this image, together with the measurements of the I-band image taken from
the ING Archive which is almost similarly deep. The ellipticity behavior reveals
central rise and a peak near the position of the nuclear ring; the
isophotes in the radius range of $12\arcsec - 50\arcsec$ (please note
that $R_{25}=52\arcsec$) are indeed round. However the most interesting
things are seen at $R>R_{25}$: the ellipticity rises to the mean value of
0.15, and the major axis position angle can be measured quite certainly
at $<PA_0>=112\degr$. We cannot be sure that we see a round outer stellar
disk inclined by $\sim 30\degr$ to the line of sight because the kinematical
parameters of the orientation of the more inner {\it gaseous} disk,
$PA_0=128\degr$  and $i=9\degr$, does not coincide with the photometric
parameters found for the outermost part of the broad-band image;
the hypothesis which may be more plausible is that the outer disk is
intrinsically oval. Unfortunately, we have no detailed kinematical
measurements at such large radii.

We have tried to decompose the whole V-image into separate photometric
subcomponents, such as an outer exponential disk and some more inner
components. Two methods were applied: the software GIDRA \citep{mrk315}
which uses 2D surface brightness modelling under constant orientation
parameters  over all the image
and iterative 1D brightness profile fitting starting from the outermost
component with subsequent subtraction of the 2D model components from
the original image; the latter method allows to vary orientation parameters
from one component to another according to isophote analysis results.
The GIDRA analysis of the V-image, under the fixed kinematical parameters
of the orientation, $PA(line-of-nodes)=128\degr$ and $i=9\degr$, with an
approximation of the seeing FWHM by $1\farcs 7$, has given TWO exponential
disks superposed, with the scalelengths of $17\farcs 6$ and $7\farcs 2$
and the central surface brightnesses, $\mu _{0,V}$, of 20.6 and 18.2
mag per square arcsecond. The third photometric component seen only in
the very center may be a de Vaucouleurs' bulge with $r_e=4\farcs 2$.
One-dimensional brightness profile fitting made with $PA_0=112\degr$
and with isophote axis ratio of $b/a=0.85$ for the outer component and
with $PA_0=13\degr$ and $b/a=0.93$ for the inner components has also given
two exponential disks: the outermost one being approximated in the radius
range of $50\arcsec - 93\arcsec$ has $\mu _{0,V}=21.04$ and $r_0=20\arcsec$
and the inner one, seen in the radius range of $15\arcsec - 42\arcsec$
after subtracting the outer disk, has $\mu _{0,V}=18.45$ and $r_0=7\farcs 2$
-- see the Fig.~\ref{disk7742}. The most central component which is left after
subtraction of two exponential disks gives a noticeable contribution only
inside $R\approx 5\arcsec$, and since it is affected by spatial resolution
effects we cannot surely determine shape of its profile: it may be exponential
as well as something else. However both our fitting methods indicate
certainly the presence of two exponential disks with different scalelengths.
We would like to stress that it is the outer disk which is `normal':
its $\mu _{0,V}$, 21 V-mag per square arcsec, is very close to the canonical
Freeman's value \citep{fr70}, and the relation between its central
surface brightness and its scalelength in kpc is typical for Sb-galaxy
\citep{dejong3}. The inner disk is more compact and high-surface-brightness
one than spiral galaxies have usually, though not so compact and
bright as circumnuclear disks of early-type galaxies; on the diagram
`$\mu _0$ vs h' collected by \citet{resh} it settles among the large disks
of lenticular galaxies. However the spiral arms and noticeable star formation
(H$\alpha$ emission) in NGC 7742 are confined just to this inner disk.

Interestingly, NGC 7217 -- another galaxy with rings and without a bar --
has very similar structure. We have decomposed its brightness profile
in our work \citep{we2000} and have found two exponential
disks, the outermost disk being the `normal' one, together with the
compact exponential bulge. We compare the structural characteristics
of the components for both galaxies in Table 2. The scalelengths are
very close: 2--3 kpc for the outer disks and $\sim 1$ kpc for the inner
disks. The visible shapes of the outer and inner disks are different
in both galaxies. However, if to compare the derived photometric
characteristics of the disks with the orientation parameters estimated
from the global gas kinematics, we would conclude that in NGC 7217 the outer
disk is round and the inner disk is inclined or is oval (a destroyed bar?),
whereas in NGC 7742 the configuration is opposite: the inner disk is
round and the outer one is oval that may be due perhaps
to an external tidal perturbation.

Recently \citet{n278} have studied another
unbarred ringed galaxy, NGC 278, having the morphological type close to
that of NGC 7217 and NGC 7742, SAB(rs)b. Their graphic presentation of the
surface brightness profile of NGC 278 allows to suggest the same multi-tiers
structure of the global stellar disk as we have found in NGC 7217 and
NGC 7742. All the present star formation
in NGC 278 is confined to the inner disk, within the radius of 1.1 kpc,
as well as in NGC 7742.

\begin{table*}
\scriptsize
\caption[ ] {Exponential parameters of the brightness profiles fitting}
\begin{flushleft}
\begin{tabular}{lcccccc}
\hline
Disk & Radius range of fitting, arcsec & Radius range of fitting, kpc
 & $PA_0$  & $b/a$ &
$r_0$, arcsec & $r_0$, kpc \\
\hline
\multicolumn{7}{l}{NGC 7217}\\
Outer & 60--110 & 5--9  & $90\degr$ & 0.82 &  35.8
& 2.9 \\
Inner & 20--50 & 1.6--4  & -- & 0.92 &
12.5 & 1.0 \\
Central bulge & 5--20 & 0.4--1.6 &  $82\degr$ & 0.88 &
3.9 & 0.3 \\
\hline
\multicolumn{7}{l}{NGC 7742}\\
Outer & 50--93 & 6--11 & $112\degr$ & 0.85 &  20
& 2.34 \\
Inner & 15--42 & 1.8--4.9 & -- & 0.93 & 7.2
& 0.8 \\
Central (bulge?) & 1--5 & 0.1--0.6 & -- & 0.93 & 1.3 & 0.15\\
\hline
\end{tabular}
\end{flushleft}
\end{table*}

Some words about the central component of NGC 7742. Its rather high visible
ellipticity was noted earlier, e.g. by \citet{n7742rep} and by
\citet{3bphot}; the former authors mentioned the turn of the isophote major
axis from $110^{\circ}\pm 10^{\circ}$ to $10^{\circ}\pm 10^{\circ}$ between
$r=1\farcs 5$ and $r=5\farcs 1$. We confirm this result and point out
that the rather high ellipticity of the isophotes at some distinct radii,
namely, at $r\approx 1\arcsec$ and at $r\approx 7\arcsec$ (Fig.~\ref{iso7742}),
makes the estimate of the major axis turn quite sure. May be anyone of these
elongated structures a bar or a compact triaxial bulge? If such triaxiality
exists in the center of NGC 7742, it would cause a Z-shaped disturbance of the
gas velocity field; and as we have seen in the previous Section~3,
the orientation of the kinematical major axis of the gas rotation,
$PA_{kin,gas}=128^{\circ}$, stays firmly between $r=2\arcsec$ and
$r\approx 40\arcsec$. So we don't see any signatures of the triaxial potential
in the center of NGC 7742. Instead we may suggest a strong warp of the
rotation and symmetry planes in the center of the galaxy: immediately inside
$R\approx 5\arcsec$ the gas rotation plane conserves the line of nodes
of the outer gaseous disk but probably increases its inclination
that may be deduced from the visible fast rotation, and closer to the center,
at $R< 1\farcs 5$, the kinematical major axis of the gas `switches'
to the `orthogonal' orientation \citep{fathithes}.

\section{The stellar kinematics and structure of NGC 7217}

As we have noted above, two galaxies with rings and without bars --
NGC 7217 and NGC 7742 -- have very similar global structures. As for
the fine features in their centers, in this work we suggest a strong warp
of the rotation plane both for the stars and for the ionized gas in the
center of NGC 7742. In NGC 7217 which is slightly less face-on we have
found a circumnuclear polar gaseous ring with the radius of
$3\arcsec -4\arcsec$ \citep{we2000}. As for the stellar
kinematics in the center of NGC 7217, from our MPFS observations, with our
spatial resolution $\sim 2\arcsec$ we have not
found any deviations from an axisymmetric rotation around the main
symmetry axis of the galaxy at $R\ge 2\arcsec$.
However, the surface distribution of the stellar velocity dispersion in the
center of NGC 7217 looked very strange, with the off-centered minimum,
and we \citep{we2000} were not able to give a reasonable explanation of it.

Now we have in hand the 2D spectral data for NGC 7217 obtained with the
SAURON; these data provide a larger field of view and a slightly better
spatial resolution than the MPFS ones so now we can expand our previous
analysis of the kinematics of the central part in this galaxy.
By applying a tilted-ring method to the whole stellar velocity field
(Fig.~\ref{sau7217}{\it right}) representing a combination of two different
pointings of the telescope, outside $R=3\arcsec$ we obtain the mean parameters
of the rotation plane orientation,
$PA_{0,kin,*}=268\degr \pm 2\degr$ and $i=30\degr \pm 4\degr$,
very stable along the radius, consistent with the axisymmetric rotation
in the main galactic plane. The velocity field of the ionized gas
(Fig.~\ref{sau7217}{\it left})
obtained by measuring the emission line [\ion{O}{3}]$\lambda$5007, strong
in the center of NGC 7217, confirms the orientation of the kinematical major
axis for the ionized gas found by us earlier \citep{we2000}:
$PA_{0,kin,gas}=329\degr \pm 4\degr$ at $R=1\arcsec - 4\arcsec$.
The angular rotation velocity is rather high,
$\omega \sin i_{gas} \approx 29$ km/s/arcsec; being compared to the
stellar rotation velocity at the same radius,
$\omega \sin i_* \approx 14.5$ km/s/arcsec, and taking into account that
$\sin i_*=0.5$, it implies the presence of the circumnuclear edge-on gaseous
disk. The higher spatial resolution of the SAURON data with respect to
the previous MPFS ones allows to notice a turn of the stellar kinematical
major axis inside $R\le 2\arcsec$, and it is a quite new finding.
Inside this radius the stellar kinematical major axis turns
and reaches $PA_{0,kin,*}=309\degr \pm 12\degr$ at $R=1\arcsec$ --
compare to $PA_{0,kin,gas}=329\degr \pm 4\degr$
for the ionized gas inside $R=3\arcsec$. There is a
clear impression that a {\it stellar} inclined disk exists too but it is
much more compact than the gaseous one. We have collected all the available
velocity fields for the central part of NGC 7217 by adding to the data
analysed in this work the Fabry-Perot ionized-gas velocity field presented
by us earlier \citep{we97c} and the CO velocity field presented
by \citet{nuga7217} recently in the frame of the NUGA project.
We have applied the tilted-ring analysis to all of them and have traced
the kinematical major axis orientation from the very center to
$R\approx 30\arcsec$. In Fig.~\ref{majax7217} we compare these
results to the photometric major axis orientation. Outside the nuclear
ring, at $R>10\arcsec$, both the gas -- warm and cold -- and stars
rotate quite axisymmetrically, with their kinematical major axes
agreeing perfectly with the photometric major axis. It is somewhat
strange because earlier \citep{we2000} we supposed the inner disk seen
at $R>20\arcsec$ to be oval because its photometric major axis, after
subtracting the other structural components, deviated by some $30\degr$
from the line of nodes of the outer disk. The only
tentative signature of possible non-circular motions
of the gas within the inner disk of NGC 7217 may be small radial
velocities, of 5--7 km/s, detected by us in the CO velocity field;
but this presence of radial gas motions is not confirmed by the results
of our analysis of the IFP ionized-gas velocity field. We must note here
that the IFP velocity measurements for the ionized gas of NGC 7217 were
made with the weak emission line [\ion{N}{2}]$\lambda$6583 and are rather
noisy and patchy so perhaps we were not able to detect radial motions
of less than 10 km/s. As for the stars, the inner stellar disk of
NGC 7217 is not very cold \citep{we2000}, so the stars may be more stable
against the weak triaxial perturbation. Inside the inner ring,
at $R\le 8\arcsec$, the kinematical major axis of the ionized gas
starts to turn implying an appearance of the inclined disk. At $R=4\arcsec$
the $PA_{0,kin,gas}\approx 320\degr$ that differs by $80\degr$ from the
direction of the photometric major axis which is here at
$PA_{phot}\approx 240\degr$. The kinematical major axis of stars starts
to turn much closer to the center, at $R<3\arcsec$, and at $R=1\arcsec$
it reaches almost the same orientation as that of the ionized gas. So
in NGC 7217 we have two circumnuclear inclined disks, gaseous and stellar,
fairly coplanar to each other, but the latter is much more compact than
the former.

Figure~\ref{sigma7217}{\it left} presents the map of the stellar velocity
dispersion in the center of NGC 7217 obtained from the SAURON data. Now, with
the larger field of view, we can unambiguously recognize the central structure
seen in this map: though the whole distribution is slightly asymmetric,
in the very center there is a certain $\sigma _*$ minimum, perhaps,
shifted by $1\arcsec -2\arcsec$ to the north. Since the color distribution
derived by us from the HST/WFPC2 data is also asymmetric, the color
maximum being shifted in the opposite direction with respect to the
stellar velocity dispersion minimum, we suggest that this asymmetry
may be caused by the dust in the inclined circumnuclear disk.  We conclude
that the stellar velocity dispersion distribution is another signature
of the compact circumnuclear stellar disk in this galaxy which must be a
relatively `cold' dynamical component.

\section{Conclusions and Discussion}

By using a variety of 2D kinematical data as well as deep images of
NGC 7742, we analyse stellar and gaseous kinematics in this unbarred
Sb galaxy possessing the nuclear star-forming ring; we compare it to
NGC 7217, another unbarred spiral galaxy with rings which has been
studied by us earlier. We have found some common features in NGC 7217
and NGC 7742.

\begin{enumerate}
\item{
Both galaxies demonstrate global structure consisting of
two exponential stellar disks with different scalelengths; the outer
disks look quite normal whereas the inner disks are compact, with
$r_0\approx 1$ kpc, and have unusual high surface brightness.

We would like to propose the following qualitative scenario to form such a
`multi-tiers' stellar disk. A few Gyrs ago there may be a sudden global gas
redistribution in the disk, due perhaps to external tidal perturbation or
minor merger. Before that event stars should form in the disk with a large,
normal scalelength, and after that when all the gas had been dropped closer
to the center the star formation should continue in the disk with a smaller
scalelength and higher surface density.}

\item{
Both galaxies, NGC 7742 and NGC 7217, have circumnuclear gaseous
disks with the radius of some 300 pc, highly inclined to the global disk
planes; the outer gas disks are, on the contrary, close to the main galactic
symmetry planes. Both galaxies possess also some counterrotationg subsystems.
NGC 7742 has all its gas in counterrotation with respect to all its stars,
with exception of some newly born stellar population in the ring, while
in NGC 7217 the gas outside $R=300$ pc corotates
the bulk of stars, but there are some 30\%\ of all stars in the
inner disk that counterrotates \citep{mk94}.

Three-dimensional dynamical simulations of the self-consistent evolution
of a stellar-gaseous galactic disk unstable with respect to bar-like
perturbations presented by \citet{sec1} proposed a scenario for
the origin of circumnuclear inclined gas rings. If initially the gas
of the global disk counterrotates the stars, then drifting to the center in a
triaxial potential of a transient bar, this gas must leave the disk plane
and accumulate on orbits strongly inclined to this plane -- only these
inclined orbits remain stable for the initially counterrotating gas near the
inner Lindblad resonances. We may suggest that the gas which is now observed
as the circumnuclear strongly inclined disks in NGC 7217 and NGC 7742 has come
from the outer parts of the galaxies, and when it was there, it counterrotated
the stars.}

\end{enumerate}

As for the problem of the origin of initially counterrotating gas, it may
be solved together with the problem of the nuclear star-forming rings origin.
If we suggest past minor merger
of a dwarf gas-rich galaxy from a retrograde orbit, this event had to
supply some amount of counterrotating gas and at the same time it might
cause an oval distortion of the stellar disk of the host galaxy that in
its turn had to produce rapid radial gas re-distribution and the nuclear
star-forming ring appearance -- all the peculiar features observed in
NGC 7217 and NGC 7742.
NGC 7742 demonstrates strong vertical gas oscillations in its
counterrotating gaseous disk implying rather recent gas accretion, NGC 7217
might possess the counterrotating gas in the past, but now it is fully
reprocessed into counterrotating stars.
\citet{n278} have detected strongly peculiar
kinematics of the neutral and ionized hydrogen beyond the optical
stellar disk in the unbarred galaxy with the rings, NGC 278, though
the galaxy is morphologically regular and quite isolated; they conclude
that the galaxy has recently experienced a minor merger.
In absence of the detailed neutral-hydrogen observations well outside the
optical borders of the galaxy, one would treat NGC 278 as a twin for NGC 7217
and NGC 7742. To our opinion, NGC 278 may represent an early stage of the
evolution having followed a minor merger, with respect to two galaxies
considered in our work, and its nearest future is perhaps NGC 7742.
{\it The presence of numerous minor merger signatures in the three unbarred
galaxies with nuclear star-forming rings makes the hypothesis of tidally
induced oval distortion of the global stellar disks the most attactive
scenario for the ring origin in unbarred galaxies.}

\acknowledgements
We thank Prof. V. L. Afanasiev for supporting the Multi-Pupil Fiber
Spectrograph of the 6m telescope and for taking part in some of the
observations which data are used in this work. We are indebted to Dr.
S. Garcia-Burillo who has provided the CO velocity field of NGC 7217
in digital form. The 6m telescope is
operated under the financial support of Science Ministry of Russia
(registration number 01-43). During our data analysis
we used the Lyon-Meudon Extragalactic Database (LEDA) supplied by the
LEDA team at the CRAL-Observatoire de Lyon (France) and the NASA/IPAC
Extragalactic Database (NED) operated by the Jet Propulsion
Laboratory, California Institute of Technology under contract with
the National Aeronautics and Space Administration. The work
is partially based on the data taken from the ING Archive of the
UK Astronomy Data Centre and on observations made with the NASA/ESA
Hubble Space Telescope, obtained from the data archive at the Space
Telescope Science Institute. STScI is operated by the Association of
Universities for Research in Astronomy, Inc. under the NASA contract
NAS 5-26555. The work on the study of global structure of disk galaxies
is partly supported by the grant of the Russian Foundation for Basic
Researches number 05-02-16454.
A.V. Moiseev thanks also the Russian Science Support Foundation.

\clearpage

\figcaption[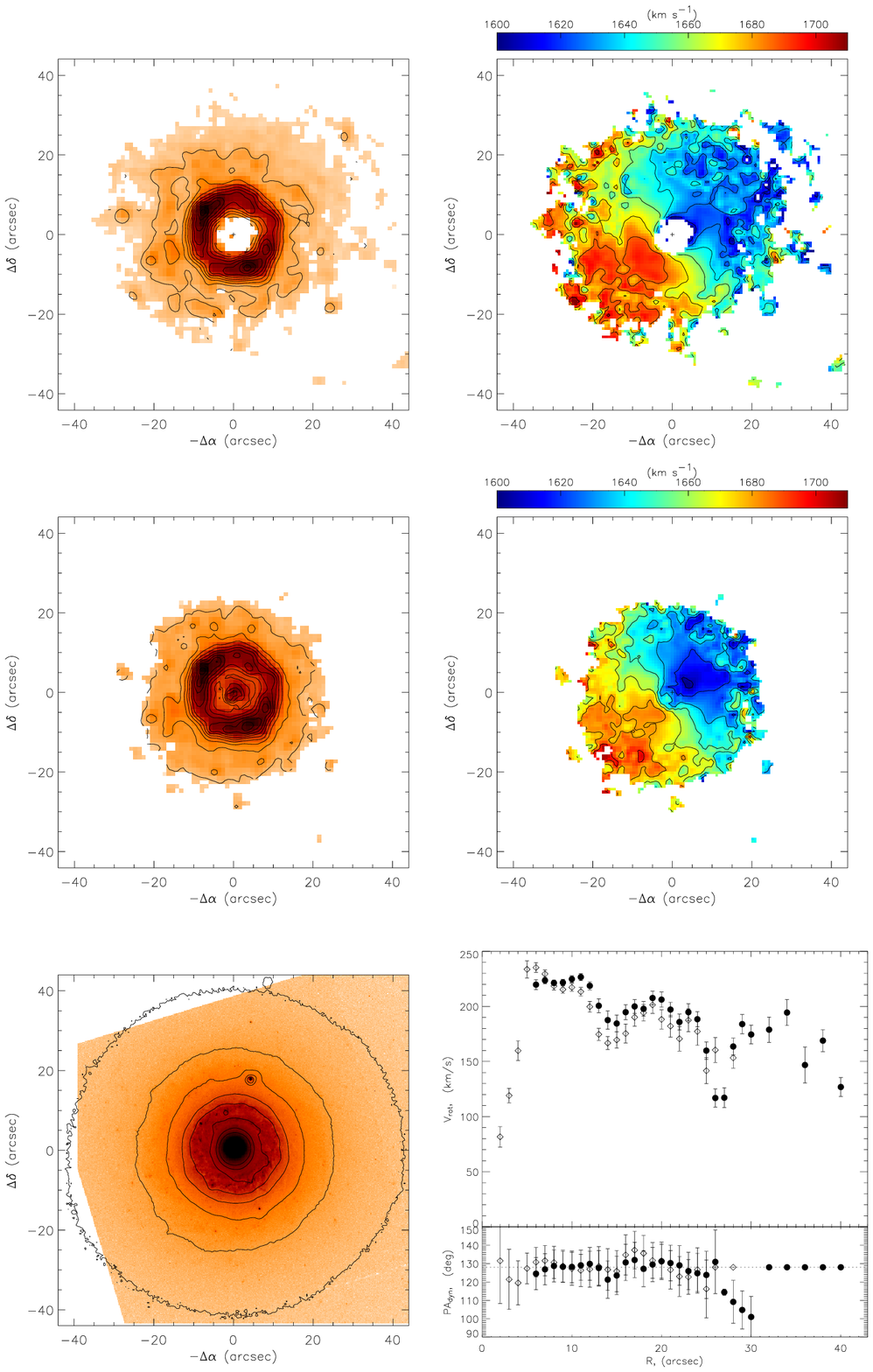]{Large-scale emission-line intensity distributions and
ionized-gas velocity fields for NGC 7742 according to our IFP data: the upper
row -- H$\alpha$, the middle row -- [\ion{N}{2}]$\lambda$6583; the bottom row
contains the continuum image, gray-scaled HST/F675W map with the broad-band
SCORPIO V-isophotes superimposed (left), and the results of the tilted-ring
analysis of the H$\alpha$ (dots) and [\ion{N}{2}] (circles) velocity fields
(right, see the text).\label{ifp7742}}

\newpage
\plotone{fig1_low.eps}
\newpage
\plotone{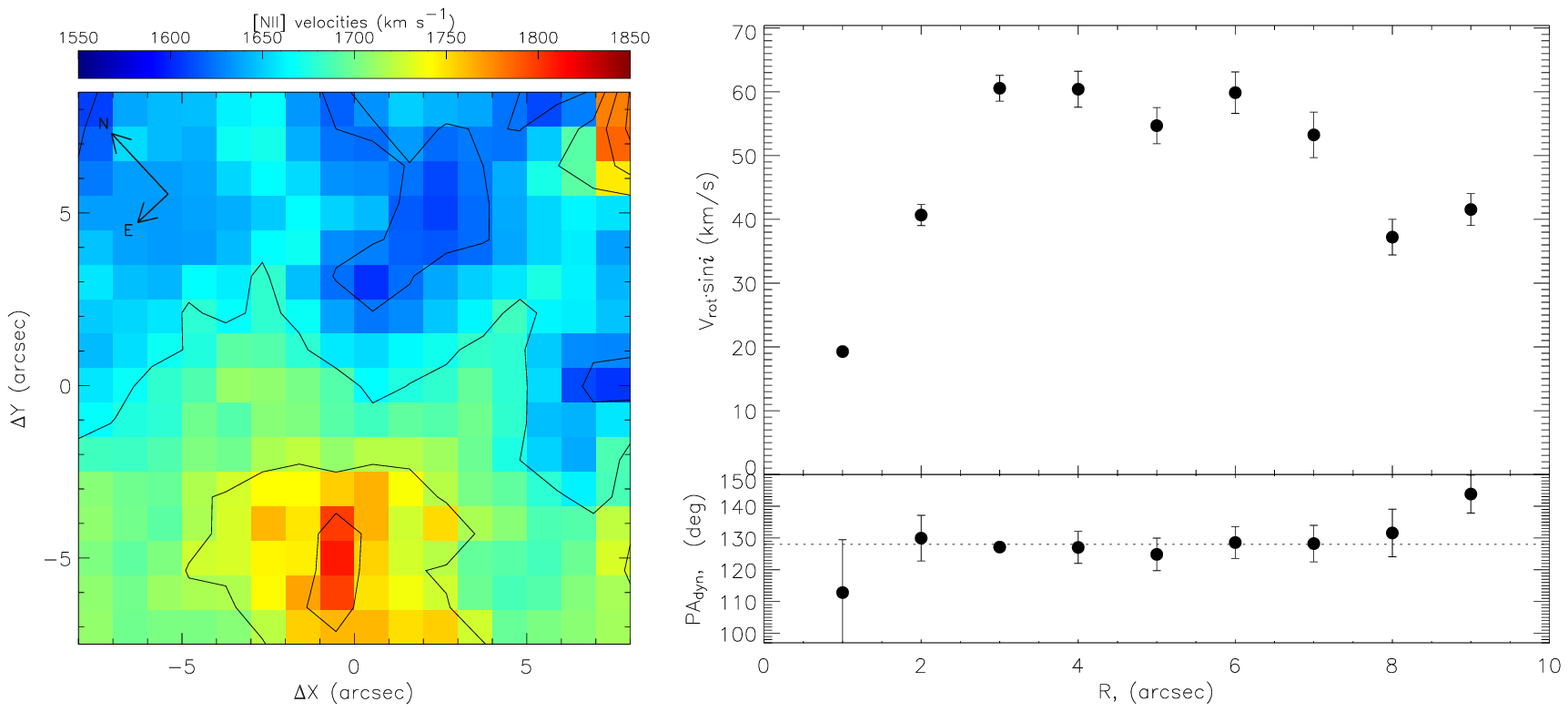}
\figcaption[fig2_col.eps]{The small-scale
[\ion{N}{2}]$\lambda$6583 velocity field of NGC 7742 according to our MPFS
data (left) and the results of its analysis (right); the inclination for the
innermost part of the gaseous disk cannot be determined
properly.\label{mpfs7742}}

\plotone{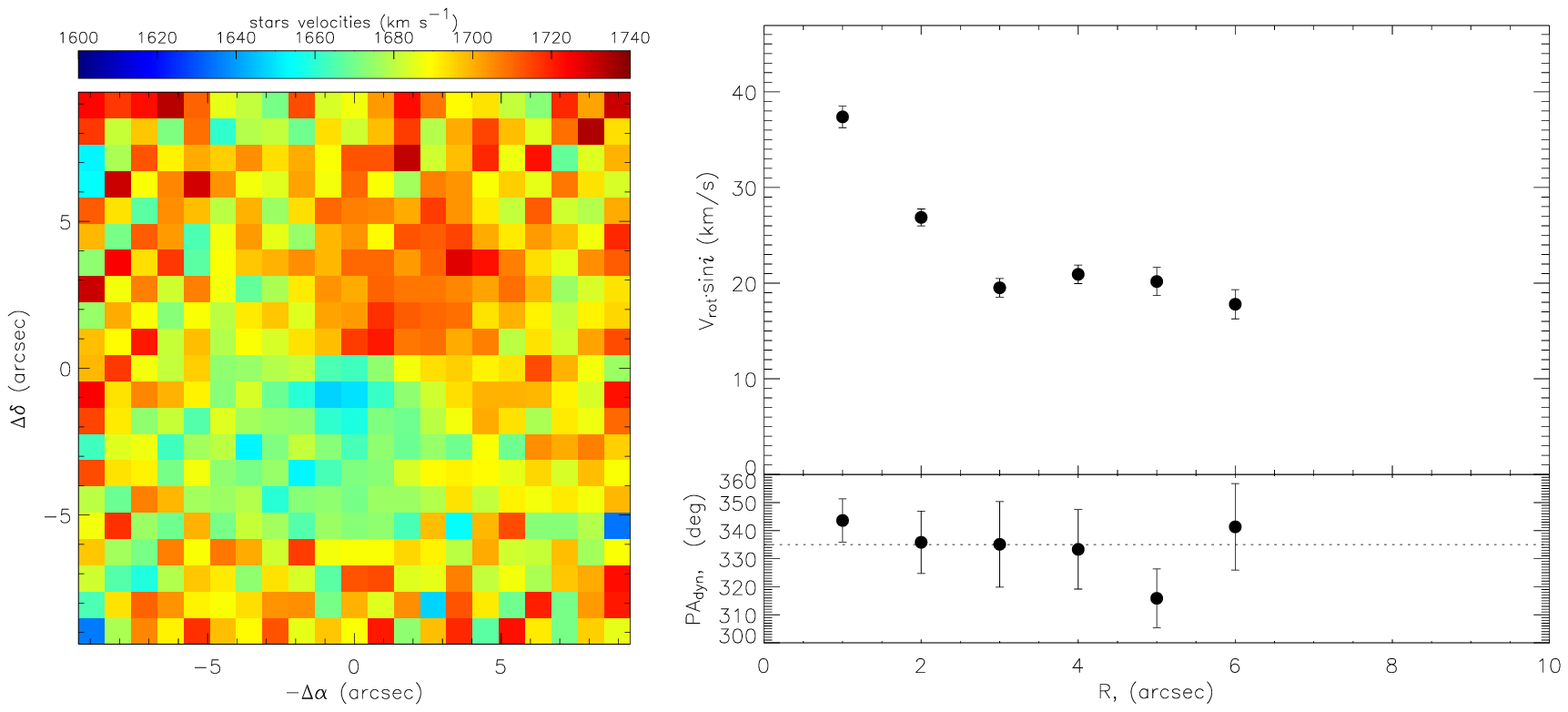}
\figcaption[fig3_col.eps]{The small-scale stellar
velocity field of NGC 7742 according to the SAURON data (left) and the results
of its analysis as concerning the azimuthally averaged projected rotation and
the kinematical major axis orientation (right); the inclination has not been
determined for the stellar rotation plane.\label{sau7742}}

\newpage
a)
\plotone{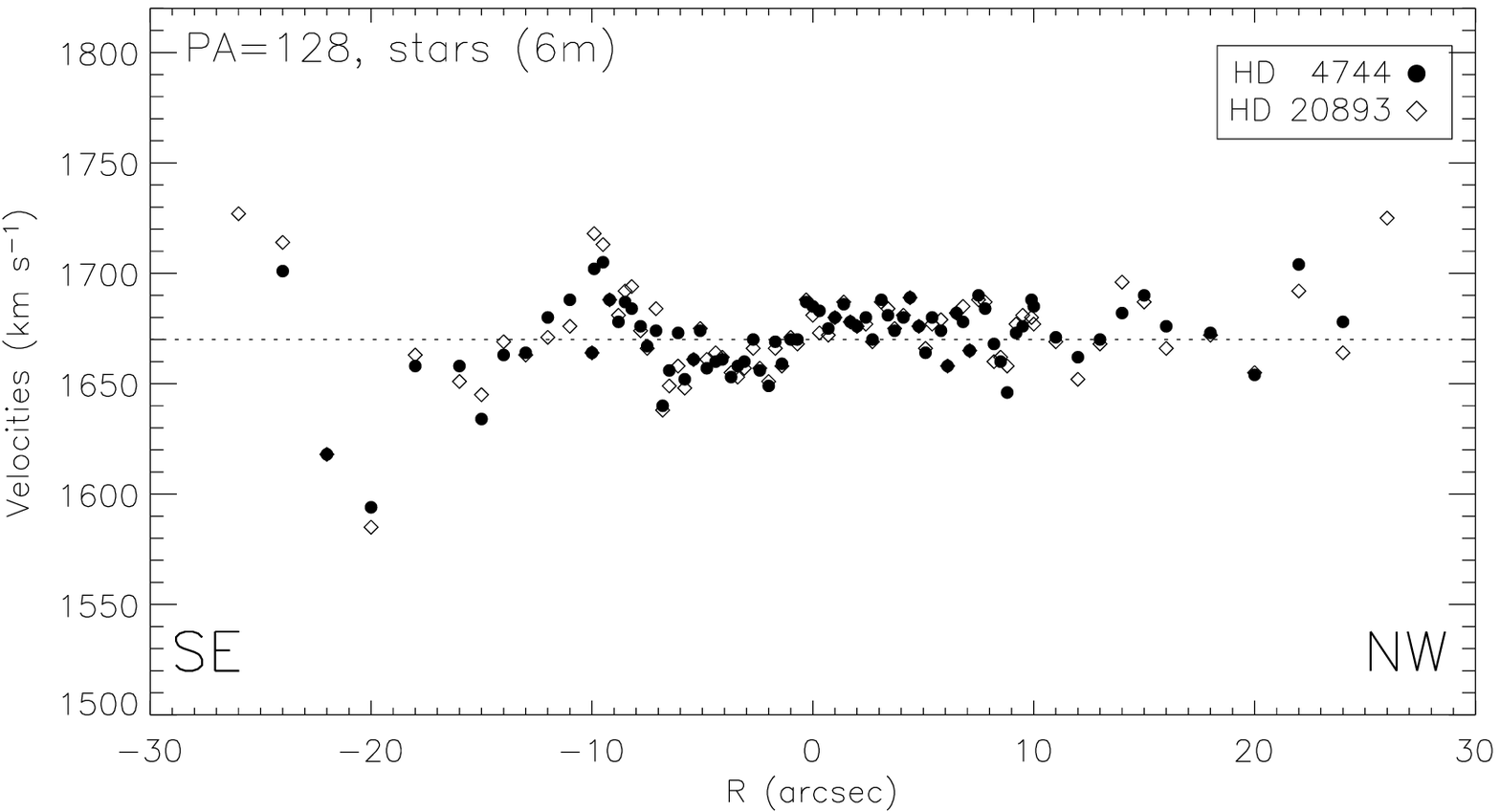}
b)
\plotone{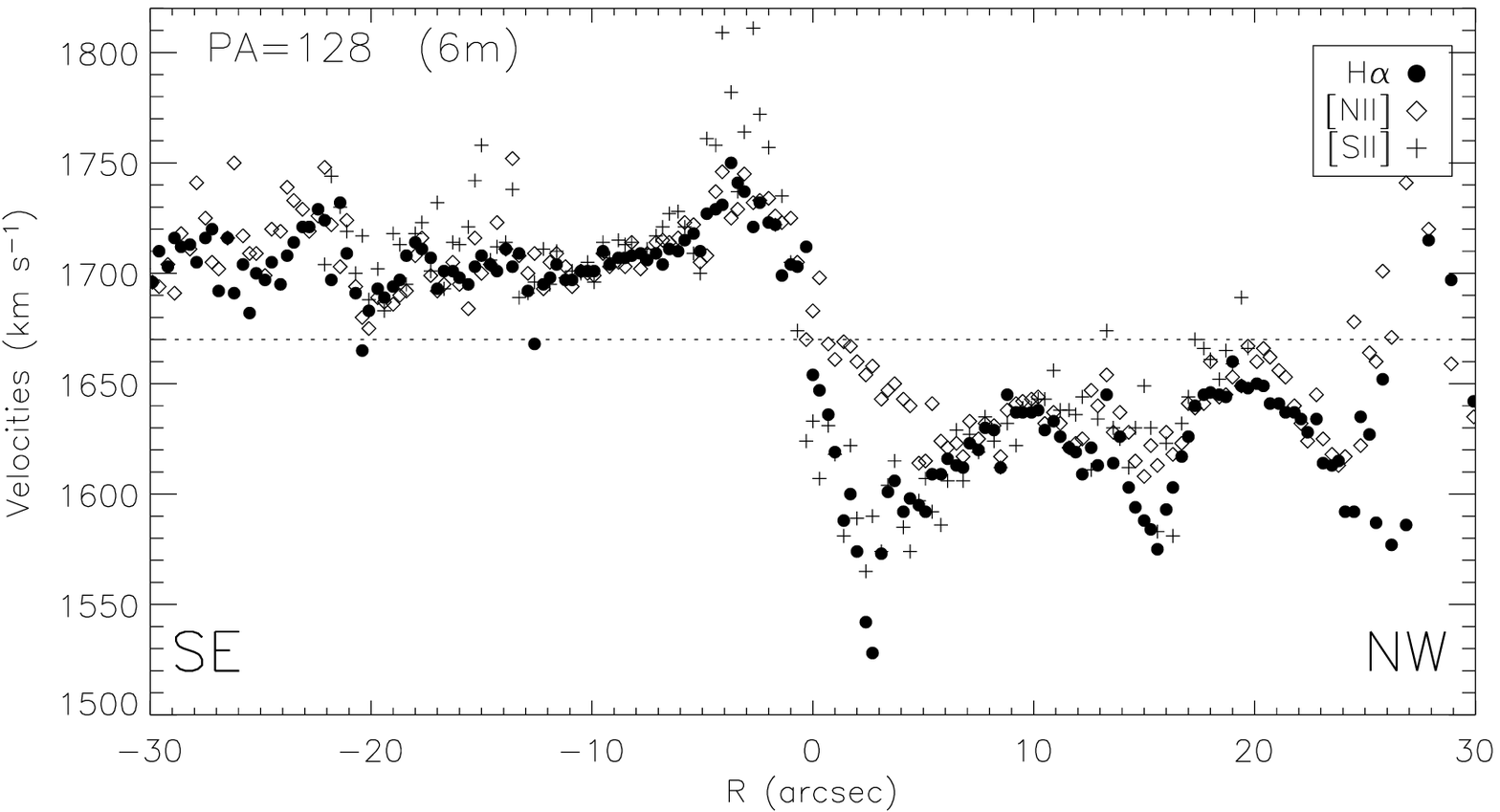}

\newpage
c)
\plotone{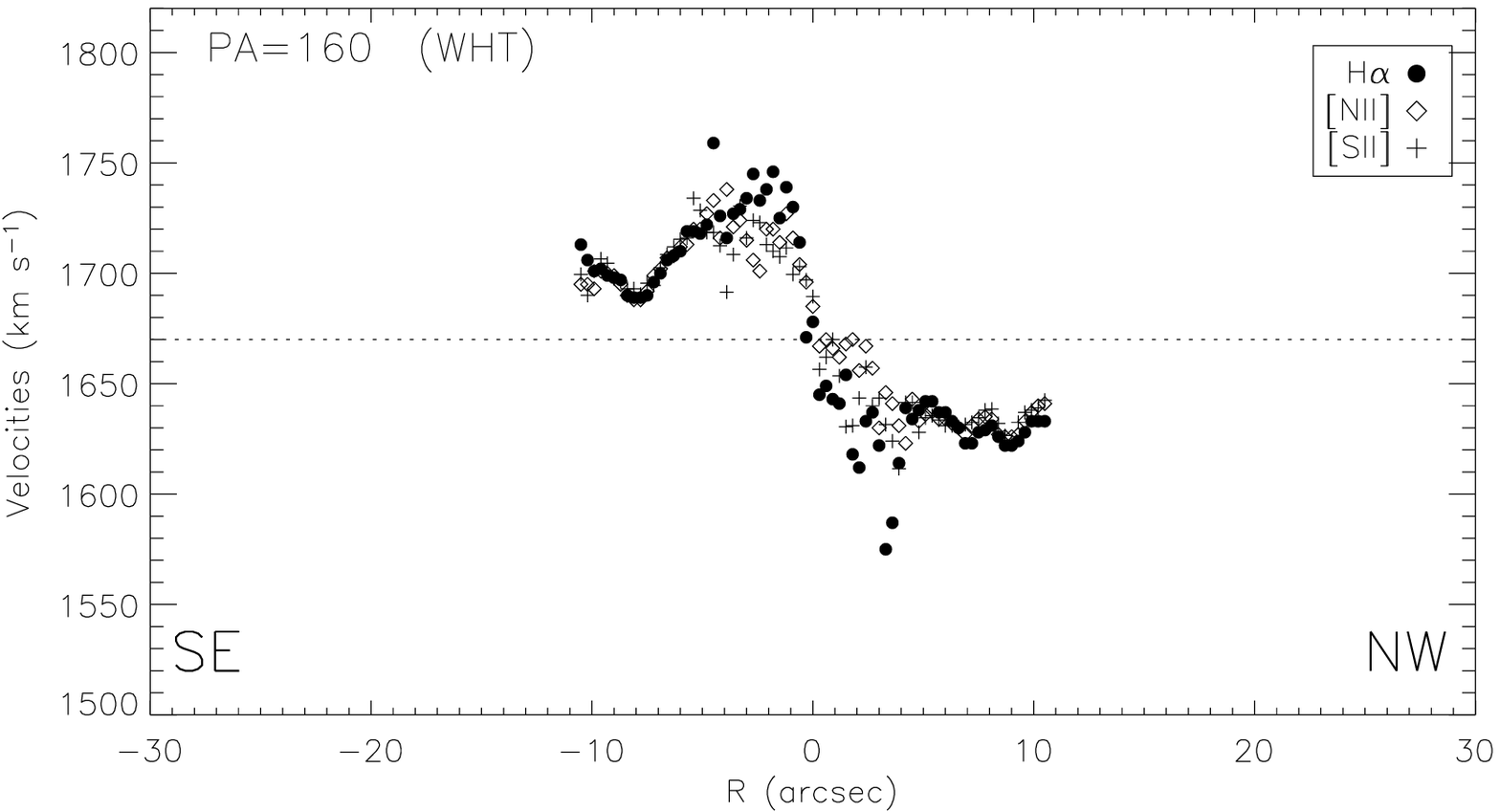}

\figcaption[f4a.ps,f4b.ps,f4c.ps]{The long-slit velocity measurements for NGC
7742: {\it a} -- stellar line-of-sight velocities according to the SCORPIO
data, two giant stars of different spectral types have been used as templates
for cross-correlation, {\it b} -- ionized-gas line-of-sight velocities from
the measurements of the various emission lines according to the SCORPIO data,
{\it c} -- ionized-gas line-of-sight velocities according to the ISIS/WHT
data.\label{ls7742}}

\newpage
\plotone{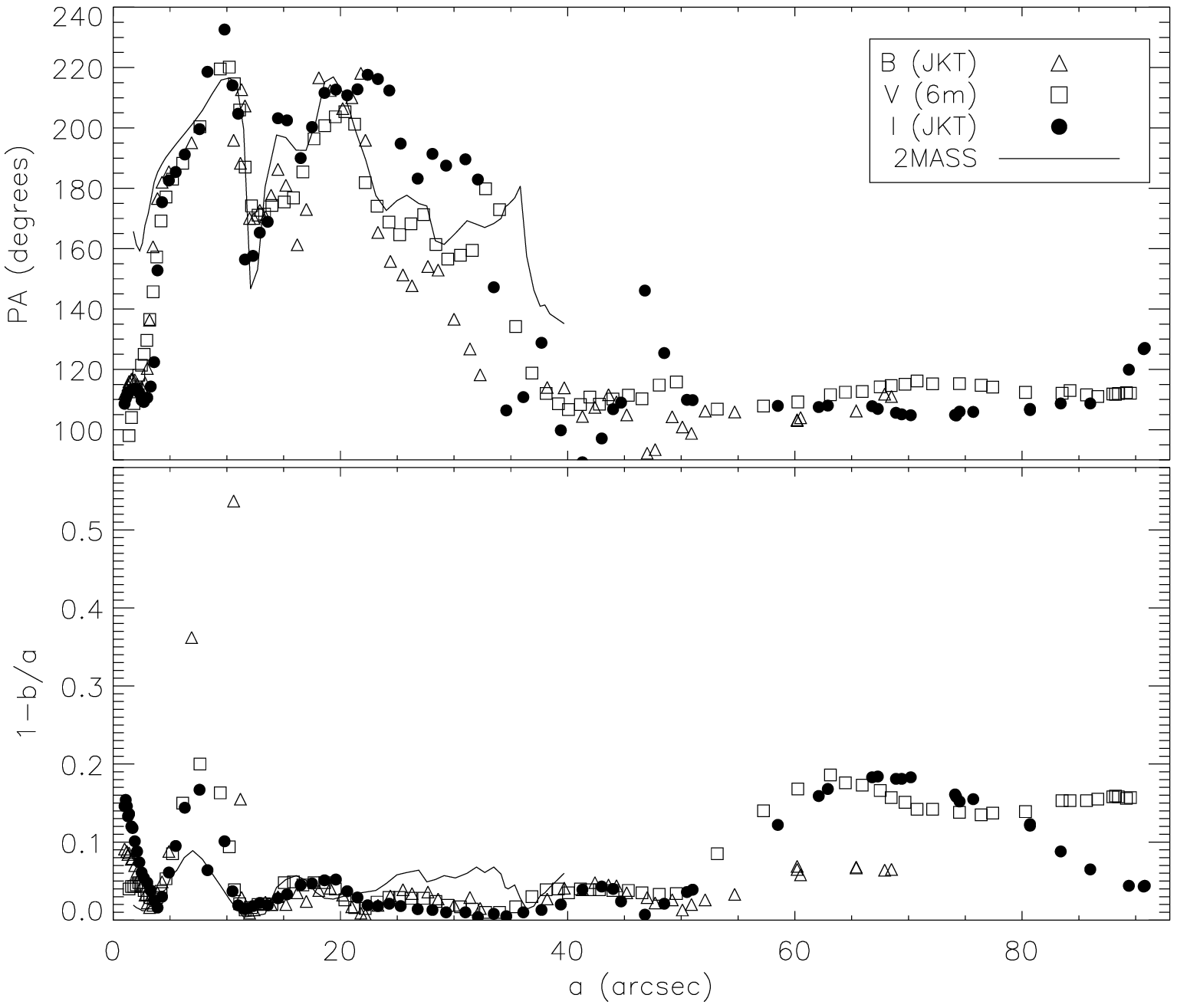}
\figcaption[f5.ps]{The results of the isophote analysis of the various
broad-band images of NGC 7742.\label{iso7742}}

\newpage
\plotone{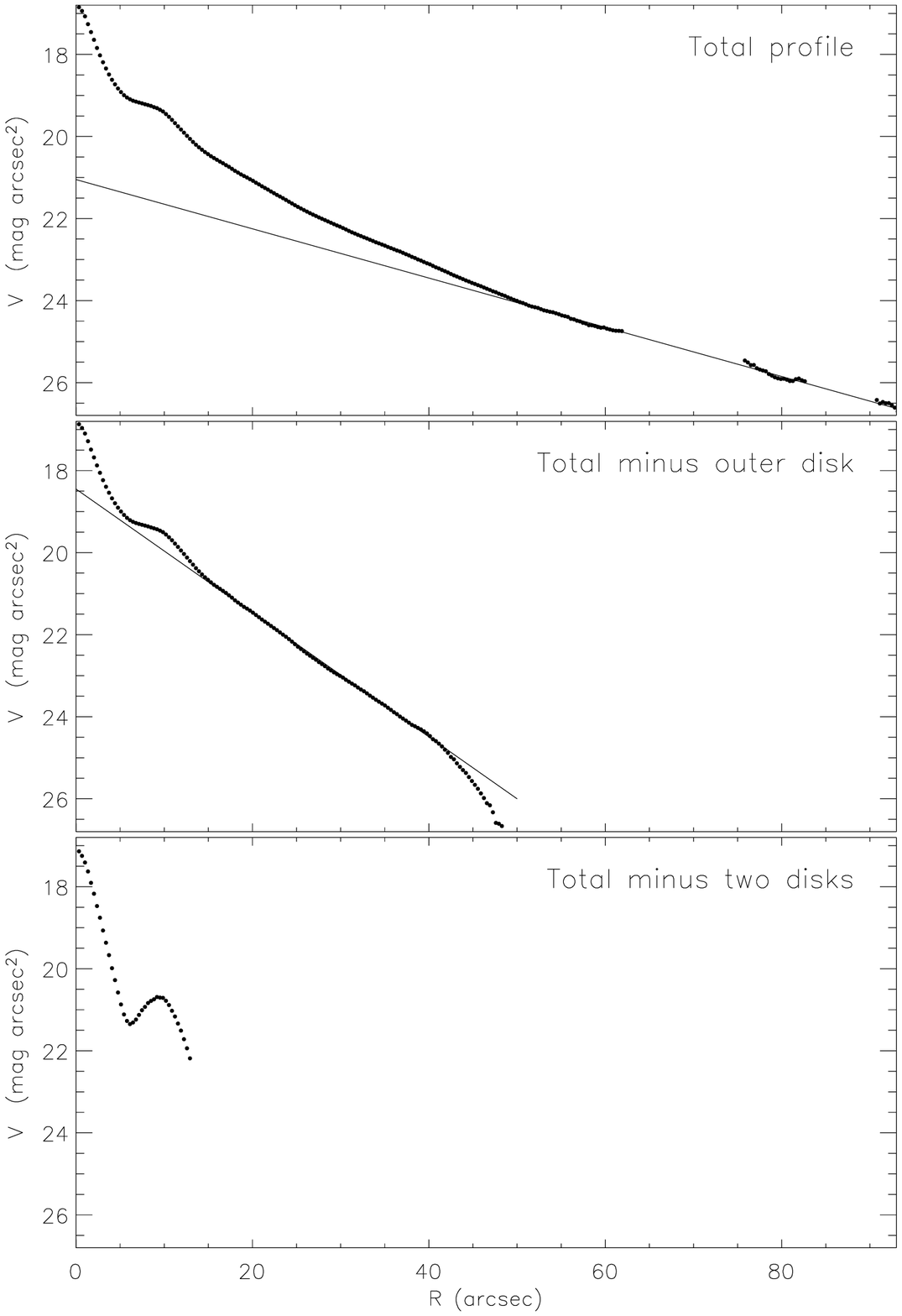}
\figcaption[f6.ps]{The results of decomposition of the
azimuthally averaged surface brightness profile calculated from the deep
V-band image (SCORPIO) of NGC 7742 into two exponential disks and
one central bulge.\label{disk7742}}

\plotone{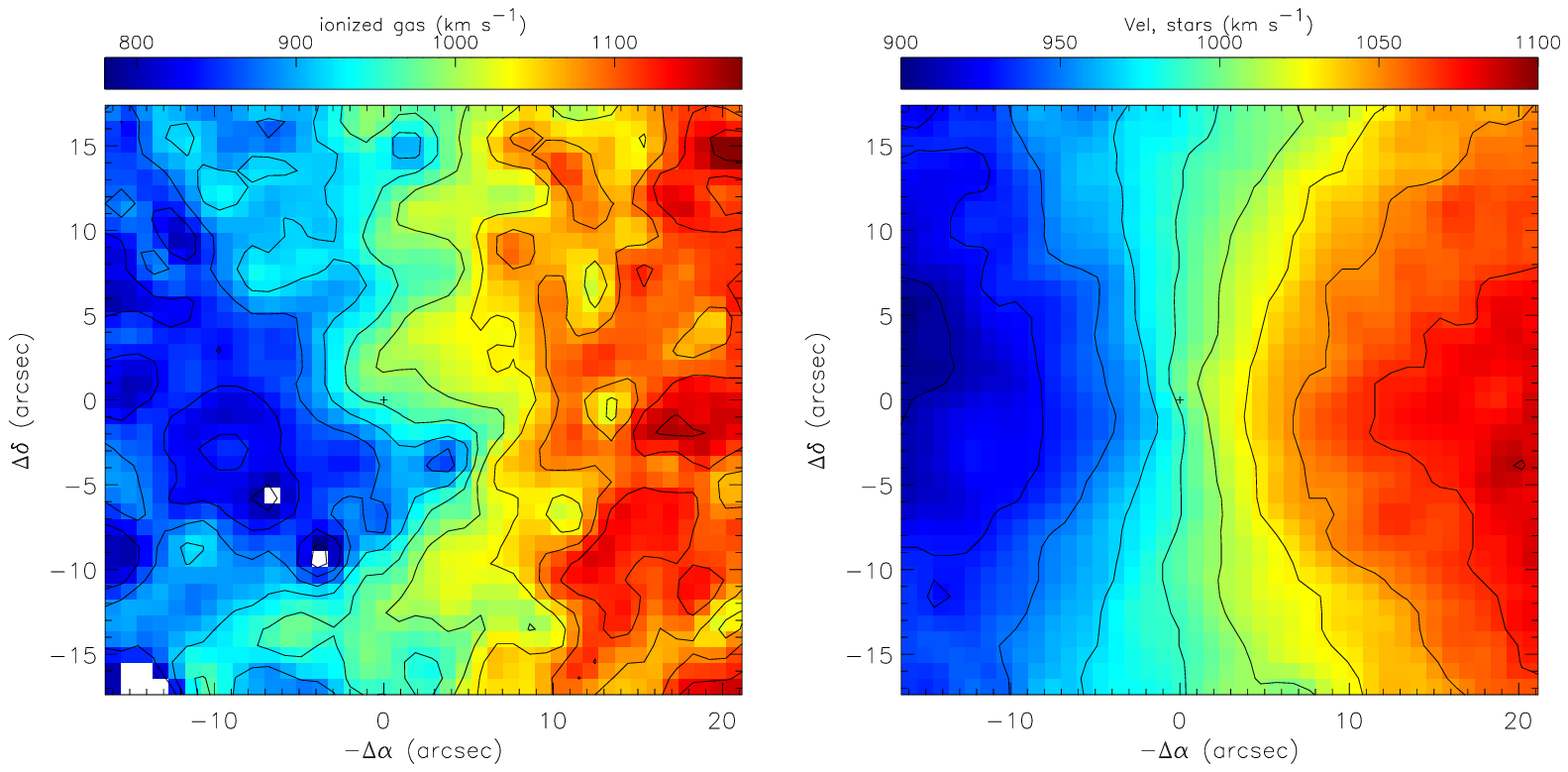}
\figcaption[fig7_col.eps]{NGC 7217: line-of-sight velocity fields for the
ionized gas (left) and for the stars (right) according to the SAURON data;
maps represent a combination of two different pointings of the WHT; the
isovelocities at the systemic velocity value are enhanced by black
colour.\label{sau7217}}

\newpage
\plotone{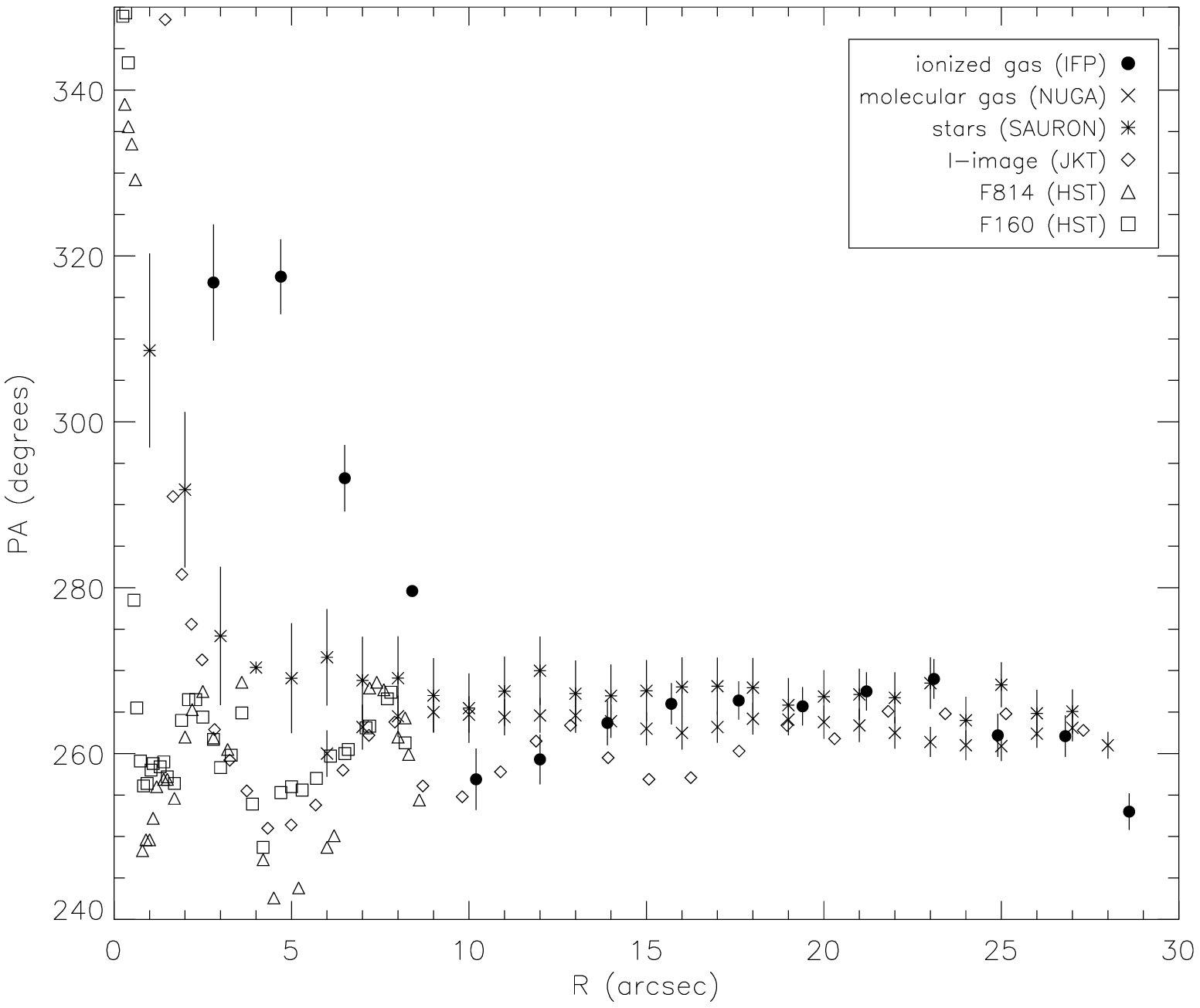}
\figcaption[f8.ps]{The comparison of the photometric (HST and JKT)
and kinematical major axis orientations in NGC 7217.\label{majax7217}}

\plotone{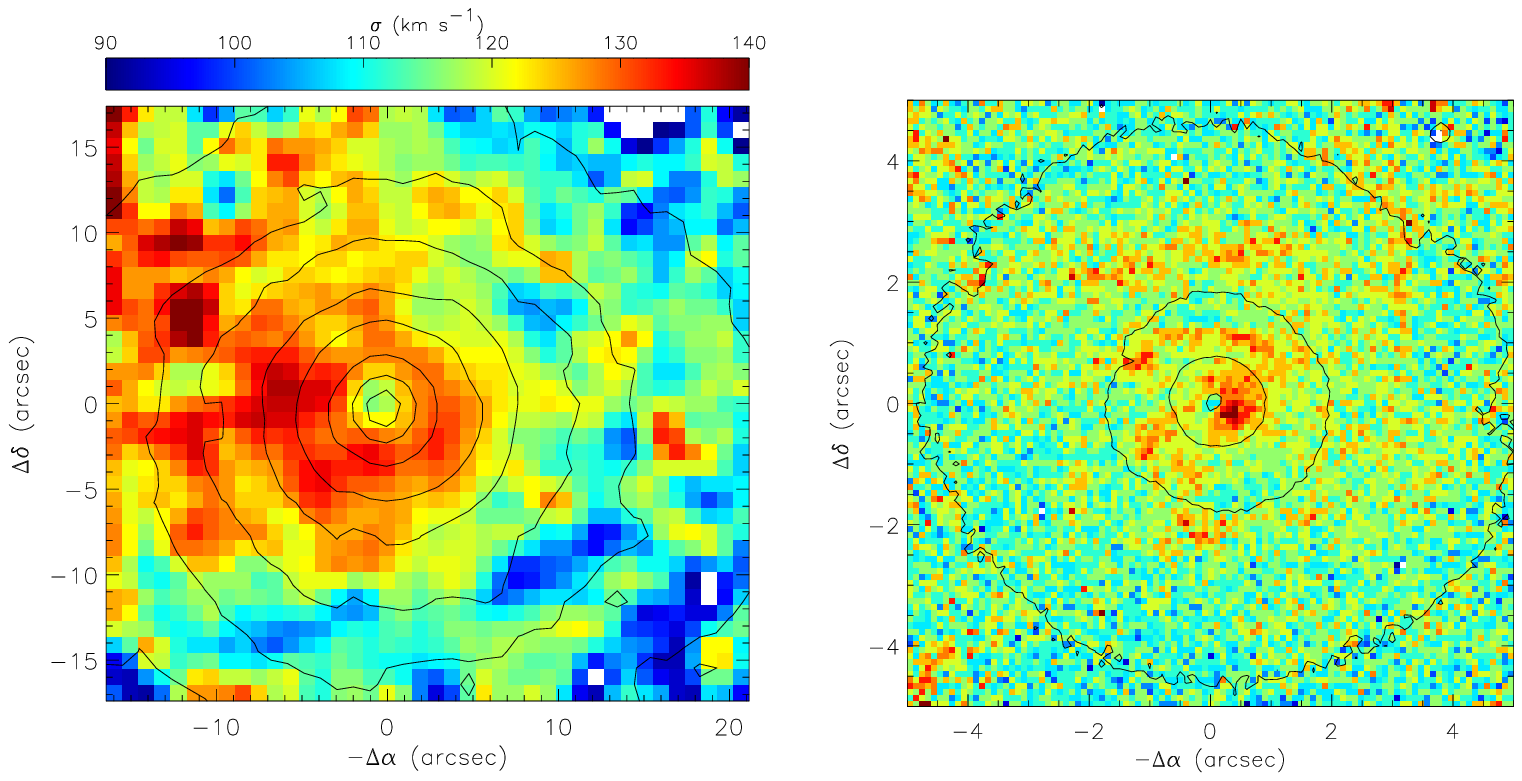}
\figcaption[fig9_col.eps]{{\it left} -- The stellar velocity dispersion map
for NGC 7217 according to the SAURON data; the continuum isophotes are
overlaid. The {\it right} plot presents the color map calculated from two
HST/WFPC2 images, F814/F606, to show the central asymmetry of the dust
distribution.\label{sigma7217}}

\end{document}